\documentclass[prb,aps,twocolumn,superscriptaddress,showpacs]{revtex4}
\usepackage{amsmath}
\usepackage{amssymb}
\usepackage{graphicx}
\usepackage{color}
\usepackage{ulem}
\usepackage[dvipdfm]{hyperref}

\begin{document}

\bibliographystyle{apsrev}

\title{Macroscopic quantum tunneling and phase diffusion in a La$_{2-x}$Sr$_x$CuO$_4$ intrinsic Josephson junction stack}


\author{Yuimaru Kubo$^{\dag\dag}$}
\altaffiliation{Present address: Quantronics Group, SPEC, CEA-Saclay, Gif-sur-Yvette 91191, France.}
\affiliation{Superconducting Materials Center, National Institute for Materials Science (NIMS), Tsukuba 305-0047, Japan}
\affiliation{Graduate School of Pure and Applied Sciences, University of Tsukuba, Tsukuba 305-8571, Japan}
\thanks{E-mail: yuimaru.kubo@cea.fr}

\author{A.O. Sboychakov$^{\dag\dag}$}
\thanks{Corresponding author. E-mail: sboycha@mail.ru}
\affiliation{Advanced Science Institute, RIKEN, Wako-shi 351-0198, Japan}
\affiliation{Institute for Theoretical and Applied Electrodynamics, Russian Academy of Sciences, Moscow 125412, Russia}
\thanks{These authors contributed equally to this work.}

%

\author{Franco Nori}
\thanks{E-mail: fnori@riken.jp}
\affiliation{Advanced Science Institute, RIKEN, Wako-shi 351-0198, Japan}
\affiliation{Department of Physics, The University of Michigan, Ann Arbor, MI 48109-1040, USA}

\author{Y.~Takahide}
\affiliation{Superconducting Materials Center, National Institute for Materials Science (NIMS), Tsukuba 305-0047, Japan}

\author{S. Ueda}
\altaffiliation{Present address: Department of Applied Physics, Tokyo University of Agriculture and Technology, Tokyo 184-8588, Japan.}
\affiliation{Superconducting Materials Center, National Institute for Materials Science (NIMS), Tsukuba 305-0047, Japan}

\author{I. Tanaka}
\author{A.T.M.N. Islam}
\altaffiliation{Present address: Helmholtz-Zentrum Berlin f\"ur Materialien und Energie GmbH, 14109 Berlin, Germany}
\affiliation{Center for Crystal Science and Technology, University of Yamanashi, Kofu 400-8511, Japan}

\author{Y. Takano}
\affiliation{Superconducting Materials Center, National Institute for Materials Science (NIMS), Tsukuba 305-0047, Japan}
\affiliation{Graduate School of Pure and Applied Sciences, University of Tsukuba, Tsukuba 305-8571, Japan}

\begin{abstract}
We performed measurements of switching current distribution in a submicron La$_{2-x}$Sr$_x$CuO$_4$ (LSCO) intrinsic Josephson junction (IJJ) stack in a wide temperature range. The escape rate saturates below approximately $2$\,K, indicating that the escape event is dominated by a  macroscopic quantum tunneling (MQT) process with a crossover temperature $T^{*}\approx2\,$K. We applied the theory of MQT for IJJ stacks, taking into account dissipation and the phase re-trapping effect in the LSCO IJJ stack. The theory is in good agreement with the experiment both in the MQT and in the thermal activation regimes.
\end{abstract}


\date{\today}

\maketitle

\section{Introduction}\label{Introduction}

Stacks of intrinsic Josephson junctions~\cite{Kleiner} (IJJs) are of considerable interest nowadays because of their possible applications in terahertz physics~\cite{Rev}, quantum electronics, etc.
Both macroscopic quantum tunneling (MQT) and thermal-activated switching in high-$T_c$ JJs, in particular in IJJ stacks of Bi$_2$Sr$_2$CaCuO$_{8+\delta}$ (BSCCO), have been reported in recent years~\cite{Inomata, Bauch, Ustinov, Li:PRL2007, Kashiwaya, Ota, Ueda, KuboAPEX}, and have attracted much interest~\cite{Mros,wePRL,weEPL,wePRB,Kawabata,Yokoyama,Koyama}. The enormous enhancement of the MQT rate~\cite{Ustinov} due to the interaction among stacked IJJs, has also attracted considerable attention~\cite{wePRL,weEPL,wePRB,Koyama}, because this is an unconventional quantum phenomenon. Switching dynamics has been studied not only in IJJs, but also in various systems, such as long  Josephson junctions of different types~\cite{Castellano,PhAT,RatchetPRL} and buckling nanobars~\cite{nanobar1,nanobar2}.

Let $T^{*}$ denote the crossover temperature distinguishing the MQT and the thermal activation regimes.
The high $T^{*}$ of IJJs (due to their high critical current density) makes them quite attractive as a candidate for a quantum bit (qubit) that can be operated at higher temperatures. From this point of view, IJJ stacks of La$_{2-x}$Sr$_x$CuO$_4$ (LSCO) compounds could be even more preferable than BSCCO stacks, since the Josephson plasma frequency in LSCO materials (and, consequently $T^{*}$) ranges from a few hundreds GHz to $\sim1$\,THz,~\cite{Tamasaku} which is higher than the typical value $\sim100$\,GHz in BSCCO. Moreover, LSCO IJJ stacks are also an interesting system to investigate the enhancement of the MQT rate~\cite{Ustinov}, because its inter-layer distance ($s = 0.7$ nm) is  shorter than that of BSCCO ($s = 1.5 $ nm), so that the interaction among the stacked IJJs is expected to be stronger.

Although the intrinsic Josephson effect in LSCO materials has been studied in several papers~\cite{Uematsu,Kim,KuboPhysicaC}, in most of them the sample sizes were not small enough for reliable MQT studies. This was due to the technical difficulty to fabricate small, submicron, LSCO IJJ stacks carved out of a bulk single crystal. Thanks to the recent progress achieved in Ref.~\onlinecite{KuboJAP}, submicron single-crystal LSCO IJJ stacks can now be obtained.

In this paper, we report precise measurements of switching currents in a submicron LSCO IJJ stack in which uniform switchings~\cite{Ustinov} occur.
The switching current distributions were measured in a wide temperature range, and the crossover temperature $T^{*}$ was determined to be $T^{*}\approx2$\,K.
In the thermal activation regime, strong phase diffusion (re-trapping) has been observed at $T\gtrsim3.5$\,K, because in LSCO the quasi-particle conductivity is high enough to suppress the escape rate.
In the phase-diffusion regime, the width of the switching histograms decreases when the temperature increases. A similar effect has been observed recently in different types of JJs.~\cite{Krasnov,TafuriNb,TafuriYBCO}

We use the theory of MQT in IJJs developed in Refs.~\onlinecite{wePRL,weEPL,wePRB}, which successfully explains the large enhancement of MQT escape rate observed~\cite{Ustinov} in BSCCO IJJ stacks.
We show that the tunneling occurs via the creation of a fluxon with a size much shorter than the junction's width, that is, the IJJ stack under study is in the {\it long}-junction regime.
This leads to the enhancement of the escape rate compared to the standard theory of MQT in {\it short} Josephson junctions.
We generalize this approach taking into account dissipation and the effect of re-trapping of the phase difference.

\section{Device and setup}\label{Device and setup}

\begin{figure}
\begin{center}
\includegraphics[width=0.48\textwidth]{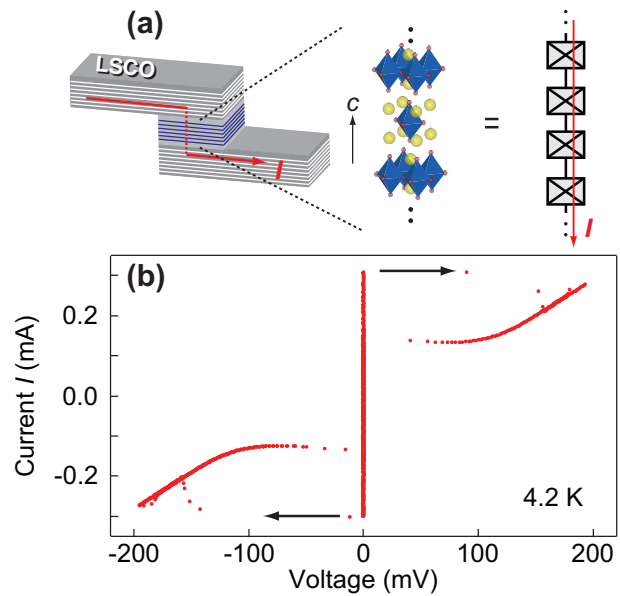}
\caption{(Color online) (a) Schematic of the geometry of the sample (left) and an array of LSCO IJJs (right). (b) A current-voltage ($I$--$V$) characteristic of the IJJ stack measured at $4.2$ K. All junctions in the stack have the same critical current, and therefore simultaneously switch to the voltage state (arrows) without showing multi-branches.}
\label{LSCOIVandIJJstack}
\end{center}
\end{figure}

A submicron LSCO IJJ stack was fabricated using the three dimensional focused-ion-beam etching technique\cite{Kim, KuboPhysicaC, KuboJAP} in a bulk La$_{1.91}$Sr$_{0.09}$CuO$_4$ single crystal grown by the traveling solvent floating zone method\cite{TanakaNature}. A schematic of the LSCO IJJ stack is depicted in Fig. \ref{LSCOIVandIJJstack}(a). The lateral dimensions of the stack are $0.45\times0.95$\,$\mu$m$^{2}$, and the thickness is about $35$\,nm (with an estimated number of layers $N\approx50$).~\cite{KuboJAP} The IJJ stack was placed in a copper box inside a $^{3}$He cryostat with a base temperature $0.4$ K. DC transport measurements were carried out with a four-terminal configuration. The low-frequency RF noise was filtered by a series of low-pass filters, and the high-frequency ($\sim$\,GHz) thermal noise was attenuated by a lossy coaxial cable placed between the $1$\,K pot and the sample box.

Switching currents of the LSCO IJJ stack were measured using a high-resolution ramp-time-based setup, which is described in detail elsewhere~\cite{Wallraff,Kashiwaya,KuboAPEX}. The bias current was linearly ramped up at a constant rate $9.17$\,mA/s, with a repetition rate $21$\,Hz. In our setup, similar to Refs.~\onlinecite{Wallraff, KuboAPEX}, the ramped bias current is turned off immediately after detecting a voltage signal above $20$\,$\mu$V from the IJJ stack, in order to minimize self-heating of the junctions. Switching current distributions $P(I)$ have been constructed from $10,000$ switching events at each temperature ranging between $0.4$ K and $10$ K.

Here we note that the $P(I)$ of a \textit{single} BSCCO IJJ having a critical current two orders of magnitude smaller ($\sim\mu$A) than that of the LSCO IJJ stack ($\sim 100\ \mu$A) was also measured using the same setup, and the results (not shown here) were in excellent agreement with the thermal activation theory for conventional single Josephson junctions.
From this measurement, the noise level of our measurement setup turned out to be much less than $\sim$10 nA.
In addition, we confirmed that the self heating is negligible at $21$ Hz, the repetition rate used in this measurement, by comparing $P(I)$'s taken at different repetition rates.
We also verified that the $P (I)$ of another sample of LSCO IJJ stack decreases under magnetic fields, as expected (see Appendix \ref{Bfield}). These results imply that any external factor does not saturate the switchings [in either $P(I)$ or $\sigma(T)$], and therefore that a reliable set of data can be obtained in our measurement setup.

\section{Switching dynamics of LSCO IJJ stack}\label{Results of the switching}

\begin{figure}
\begin{center}
\includegraphics*[width=0.48\textwidth]{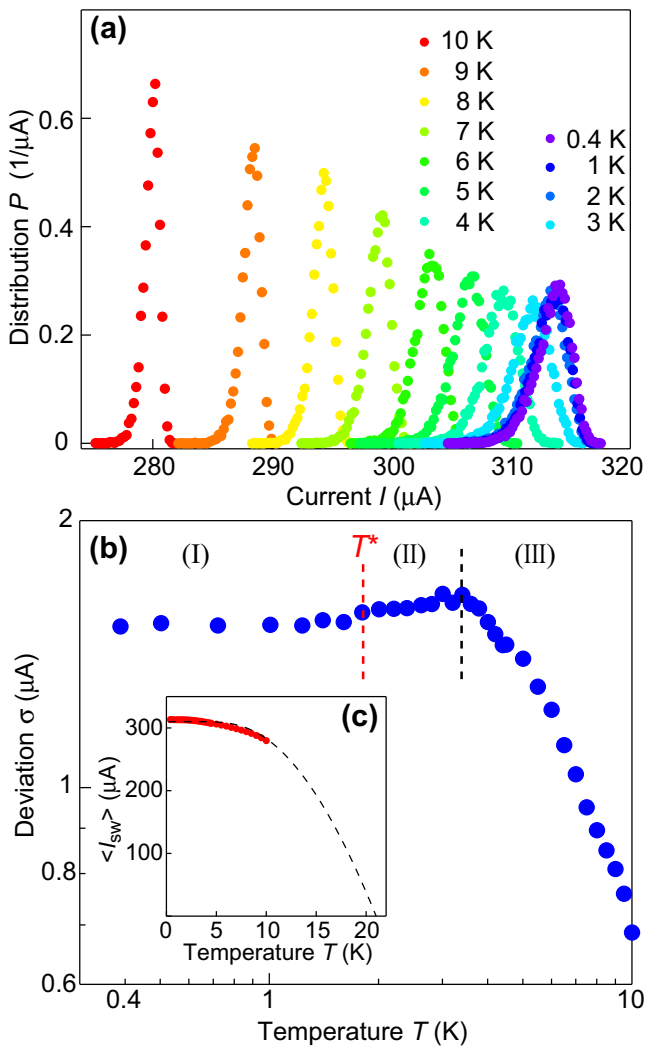}
\end{center}
\caption{\label{FigP}(Color online) (a) Switching current distributions $P(I)$ at different temperatures.
Here the bin width is 200 nA.
(b) The width (standard deviation $\sigma$) of $P(I)$ as a function of temperature $T$.
The vertical dashed lines qualitatively separate the different temperature regimes: (I) MQT, (II) cross-over, and (III) thermal activation with phase re-trapping (phase diffusion). (c) Mean switching current $<I_{\rm sw}>$ versus $T$.  The dots are experimental data, and the dashed curve is a calculation of the Ambegaokar-Baratoff theory~\cite{Ambegaokar_Baratoff} with assuming that the superconducting gap is of BCS type.}
\end{figure}

A current-voltage ($I$--$V$) characteristic of the LSCO IJJ stack is shown in Fig.~\ref{LSCOIVandIJJstack}(b).
Unlike typical $I$--$V$ curves of BSCCO,\cite{Kleiner, Inomata, Kashiwaya, Ota, Ustinov, KuboAPEX} no clear multi-branches appear, but all the junctions simultaneously switch to their voltage states.
This simultaneous switching event is due to the fact that all the stacked LSCO IJJs are homogeneous, and due to the strong coupling~\cite{Goldobin} among the IJJs.
The results of the switching measurement are shown in Fig.~\ref{FigP}(a).
The distribution $P(I)$ becomes wider when increasing the position of the peak, as the temperature decreases from $10$ K down to $\sim3.5$ K. To see this behavior of $P(I)$ more clearly, the width (standard deviation) $\sigma$ of $P(I)$ is plotted as a function of temperature $T$ in Fig.~\ref{FigP}(b). As shown in the regime (III), $\sigma$ increases with decreasing temperature: from $\sim10$ K down to $\sim3.5$ K.
Here we note that the mean switching current $<I_{sw}>$ of the LSCO IJJ stack obeys the Ambegarkar-Baratoff formula as shown in Fig. \ref{FigP}(c), indicating that our LSCO IJJ stack consists of an array of tunnel junctions.

From $3.5$\,K down to $2$\,K, the width $\sigma$ of $P(I)$ shrinks very slightly, as seen in the temperature regime (II) in Fig.~\ref{FigP}(b). Finally $P(I)$ does not change below $\sim2$\,K, and therefore $\sigma$ saturates, as seen in (I) in Fig.~\ref{FigP}(b). The saturation of $\sigma$ in the lower-temperature regime (I) qualitatively suggests that MQT occurs in these temperatures.
The temperature dependence of $\sigma$ in the regime (III), however, does not correspond to the power law $\sigma \propto T^{2/3}$, which is expected in the thermal activation theory of conventional single junctions~\cite{Wallraff}.
Thus the switching dynamics of the LSCO IJJ stack cannot be understood only by the conventional MQT and thermal activation theories for single junctions. In the following sections we quantitatively discuss those switchings' dynamics, which are in good agreement with the MQT theory for stacked IJJs~\cite{wePRL,weEPL,wePRB} and thermal activation theory in the presence of moderate damping~\cite{Krasnov,TafuriNb,TafuriYBCO,Jacob}.

\section{Theoretical model and analysis}

Since the position of the peaks in the current distribution $P(I)$ does not change at low temperatures [Fig. \ref{FigP}(a)], we assume that the switching events are dominated by MQT. The crossover temperature $T^{*}$ between MQT and thermal-activated regimes is approximately $2$ K.
For temperatures larger than approximately $3.5\,$K, $\sigma$ starts to decrease when the temperature increases.
Let us first focus in the low-temperature regime $T<T^{*}$. Afterwards we will discuss the thermal-activated regime, especially the unusual behavior of $\sigma$ with temperature, by taking into account the phase re-trapping (phase diffusion) effect due to dissipation.

\subsection{Regime of macroscopic quantum tunneling}

In the MQT regime, it is more convenient to consider the escape rate $\Gamma(I)$ instead of the switching probability $P(I)$. The escape rate $\Gamma(I)$ can be calculated from $P(I)$ by means of the formula
\begin{equation}\label{Gamma_P}
\Gamma(I)=\frac{(dI/dt)\,P(I)}{1-\int\limits_{0}^{I}dI'\,P(I')}\,,
\end{equation}
where $dI/dt$ ($= 9.17$ mA/s) is the rate of current ramp in the switching measurements. The standard theory of MQT in Josephson junctions gives the following formula for the escape rate~\cite{Caldeira}:
\begin{equation}\label{Gamma0}
\Gamma_{\mathrm{MQT}}=\frac{\omega_{p}}{2\pi}a\sqrt{120\pi B}\left[1-(I/I_c)^2\right]^{1/4}\exp(-B),
\end{equation}
where $I$ is a bias current, $I_c$ is the critical current of the junction, $\omega_{p}$ is the plasma frequency at zero bias current, and the prefactor $a$ is of the order of unity (for more details about $a$, see, e.g., the review in Ref.~\onlinecite{PreArev}). Also $B=S/\hbar$, where $S$ is the action of the system calculated for an imaginary-time trajectory of the phase difference $\varphi(\tau)$ (bounce solution). For a single short Josephson junction, and for a current $I$ close to a critical current $I_c$, the result for the tunneling exponent in the absence of dissipation is~\cite{Caldeira}
\begin{equation}\label{B0}
B=\frac{12I_c}{5e\omega_p}\frac{\left[1-(I/I_c)^2\right]^{5/4}}{(I/I_c)^2}\,.
\end{equation}
For the LSCO IJJ stack under study: $I_c=344$\,$\mu$A and $f_p=\omega_p/2\pi=440$\,GHz (see Appendix). Substituting these values into Eqs.~\eqref{Gamma0} and~\eqref{B0}, we obtain the result for $\Gamma(I)$, which is many orders of magnitude smaller than the experimental value for all currents measured. As we will show below, this is related to the fact that the junction under study is long, and the phase difference turns out to be spatially inhomogeneous. The tunneling occurs by creating a fluxon at the junction's edge with a characteristic size (in the in-plane direction)  of about $\gamma s\ll D$, where $\gamma$ is the anisotropy parameter and $s$ is the distance between CuO$_2$ superconducting layers. In addition, the phase difference is inhomogeneous in the out-of-plane direction.


\begin{figure}
\begin{center}
\includegraphics*[width=0.4\textwidth]{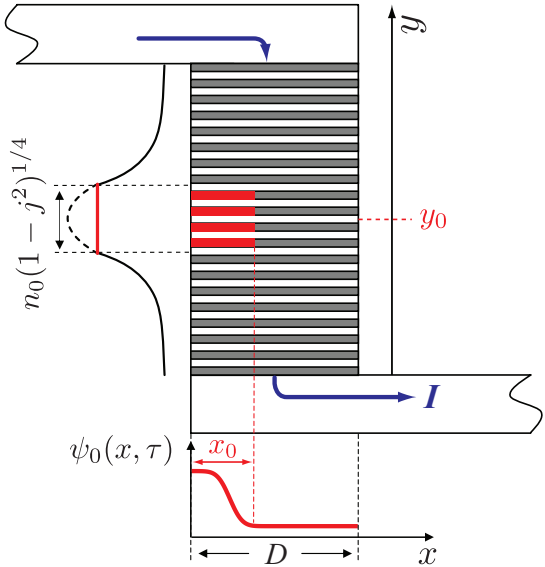}
\end{center}
\caption{\label{FigJunction}(Color online) Schematic diagram of the sample considered in the theoretical model.
The phase difference $\psi(x,y,\tau)$ is assumed to be $y$-independent in some region in the middle of the stack, and exponentially decreases outside it. The phase difference $\psi_0(x,\tau)$ is inhomogeneous in the $x$-direction, giving the characteristic size $x_{0}$ of the tunneling fluxon.}
\end{figure}

We start from the  model of coupled intrinsic Josephson junctions formed by CuO$_2$ superconducting layers~\cite{Rev}, with $\varphi_n$ being the phase difference between the $n$th and $(n+1)$th layers in the stack.
The number of stacked IJJs was estimated to be $N\approx50$ by measuring the thickness of the IJJ stack and the voltage gap in an $I$--$V$ curve. The schematic diagram of the sample is shown in Fig.~\ref{FigJunction}. Two superconducting bars (with the thickness about $3$\,$\mu$m) overlap a length $D$ in the $x$-direction.
A DC-bias current flows along the $y$-direction (perpendicular to the CuO$_{2}$ planes) inside the IJJ stack, whereas it flows along the $x$-direction in all other places (see the arrows in Fig.~\ref{FigJunction}). The $y$-axis is perpendicular to the area of the IJJ stack.
The phase differences $\varphi_n$ are assumed to be spatially inhomogeneous in the $x$-direction.

We represent $\varphi_n$ in the form $\varphi_n=\varphi_0+\psi_n(x,t)$, where $\varphi_0=\arcsin(I/I_c)$ corresponds to the equilibrium value of the phase difference.  The $\psi_n$ satisfy the equation of motion~\cite{Rev,Sakai}
\begin{eqnarray}\label{EqPsi}
&&\left(1-\frac{\lambda_{ab}^2}{s^2}\partial_n^2\right)\left[\frac{\partial^2\psi_n}{\partial t^2}+%
\alpha\frac{\partial\psi_n}{\partial t}+j(\cos\psi_n-1)\right.\nonumber\\%
&&\left.\phantom{\frac12}\!\!+\sqrt{1-j^2}\sin\psi_n\right]-\frac{\partial^2\psi_n}{\partial x^2}=0\,,
\end{eqnarray}
where $j=I/I_c$ is a normalized current, $s$ is the interlayer distance, $\lambda_{ab}$ is the in-plane penetration depth, $\alpha=4\pi\sigma_{\bot}/(\varepsilon\omega_{p})$ is the dissipation parameter, where $\varepsilon$ is the dielectric constant of the insulating layer, and $\sigma_{\bot}$ is the quasiparticle conductivity across the layers.
In Eq.~\eqref{EqPsi}, the time $t$ is normalized by $1/\omega_p$, the $x$-coordinate is normalized by the out-of-plane penetration depth $\lambda_c$, and the action of the $\partial_n^2$ operator is $\partial_n^2f_n=f_{n+1}-2f_n+f_{n-1}$.
We will use below the continuum limit, replacing $\psi_n(x,t)$ by the continuum function $\psi(x,y,t)$, where the $y$-coordinate (perpendicular to the layers) is normalized by $\lambda_{ab}$.
In this case, we can replace $(\lambda_{ab}^2/s^2)\partial_n^2\to\partial^2/\partial y^2$. Note that the inhomogeneity of the $\psi(x,y,t)$ in the $y$-direction makes the characteristic scale of the problem in the $x$-direction equal to $\gamma s$ ($\ll D$), instead of $\lambda_c$($\gg D$) if all junctions would simultaneously switch to the resistive state~\cite{Sakai} (i.e. when all $\psi_n$ are equal to each other).

To estimate the dissipation parameter $\alpha=4\pi\sigma_{\bot}/(\varepsilon\omega_{p})$, we use the Ambegaokar-Baratoff formula for the critical current~\cite{Ambegaokar_Baratoff} (which works well for LSCO~\cite{Shibauchi}), given by $I_cs/(\sigma_{\bot}A)=\pi\Delta_0/(2e)$, where $A$ is the junction's area and $\Delta_0$ is the superconducting gap at zero temperature. Using the BCS relation $\Delta_0=1.76k_BT_c$, where $k_{B}$ is the Boltzmann constant and $T_{c}$ is the superconducting transition temperature, and taking the value of the dielectric constant~\cite{Tamasaku} $\varepsilon=25$, we obtain the estimation $\alpha\sim0.2$ corresponding to moderate damping (cf. $\alpha\ll1$ for BSCCO).

To find the escape rate $\Gamma$ one should find a periodic solution to the equation of motion for $\psi(x,y,\tau)$ in imaginary time $t=i\tau$, with period $\hbar/k_BT$, and calculate the action $S$ of the system corresponding to this equation (which is actually non-local in both the $y$-coordinate and the imaginary time $\tau$).
This is a hard numerical task.
Here, we use an approximate variational approach similar to that developed in Refs.~\onlinecite{wePRL,weEPL,wePRB}.
There, it is assumed that the tunneling occurs mainly in one junction in the stack, say at point $y_0$, creating exponentially decaying tails of the phase difference outside it, that is $\psi(x,y,\tau)\propto\exp(-\kappa|y-y_0|)$ (see Fig. \ref{FigJunction}).
The analysis of  the linearized Eq.~\eqref{EqPsi} shows that $\kappa\approx\pi\lambda_c/[D(1-j^2)^{1/4}]$, where $D$ is the junction width in the $x$-direction.
For the BSCCO samples considered in Ref.~\onlinecite{wePRB}, $\kappa s/\lambda_{ab}>1$.
This means that the phase difference indeed decreases fast with increasing $y$, and one can consider the tunneling through one junction.
By contrast, we have $\kappa s/\lambda_{ab}\lesssim1$ for the LSCO IJJ stack under study. So, we should take into account a wider distribution of the phase difference $\psi(x,y,\tau)$ in the $y$-direction. It actually reflects the fact that in LSCO $s$ is smaller than that of BSCCO, so that the interaction among CuO$_{2}$ is stronger.

Here, we assume the following profile for the phase difference $\psi(x,y,\tau)$ in the $y$-direction: $\psi(x,y,\tau)=\psi_0(x,\tau)$ inside the region $|y-y_0|<\bar{n}_0=n_0(1-j^2)^{1/4}$, and $\psi(x,y,\tau)$ exponentially decays outside this region (see Fig.~\ref{FigJunction}). The parameter $n_0\sim D\lambda_{ab}/(2\pi\lambda_c s)$ can be considered as the effective number of junctions taking part in the tunneling process.
Taking the following estimates: the anisotropy parameter $\gamma=\lambda_c/\lambda_{ab}=30$, $s=0.7$\,nm, and $D=0.95$\,$\mu$m, we obtain $n_0\approx5$.
Note also the additional factor $(1-j^2)^{1/4}$, which increases the exponent $\kappa$ and makes the number of junctions taking part in the tunneling smaller for $I$ closer to $I_{c}$.

Now we obtain the effective action for the phase difference $\psi_0(x,\tau)$ in a way similar to that used in Ref.~\onlinecite{wePRB}.
We solve the linearized Eq.~\eqref{EqPsi} at $|y-y_0|>\bar{n}_0$, with the boundary condition $\psi(y_0,x,\tau)=\psi_0(x,\tau)$.
Using then Eq.~\eqref{EqPsi} and Maxwell equations, we obtain the relation for $\psi_0(x,\tau)$ and the action corresponding to it (for details, see Ref.~\onlinecite{wePRB}).
As a result, we find:
\begin{eqnarray}\label{Spsi}
S_{\text{eff}}&=&\frac{\Lambda}{d}\int\limits_0^{\tau_0}\!\!d\tau\!\!\int\limits_0^d\!\!dx\left\{\frac12\left(\frac{\partial\psi_0}{\partial\tau}\right)^2\right.
+\mu_0(1-\cos\psi_0)\nonumber\\
&-&\!\!\!j(\psi_0-\sin\psi_0)+\frac{\mu_0n_0\gamma s}{4\pi D}\frac{\partial\psi_0}{\partial x}\!\!\int\limits_0^d\!\!dx'P(x,x')\frac{\partial\psi_0}{\partial x'}\nonumber\\
&+&\!\!\!\left.\frac{\alpha}{4\pi}\int\limits_{-\infty}^{\infty}\!\!d\tau'\frac{\left[\psi_0(x,\tau)-\psi_0(x,\tau')\right]^2}{(\tau-\tau')^2}\right\},
\end{eqnarray}
where $\mu_0=\sqrt{1-j^2}$, $d=D/\lambda_c$, $\tau_0=\hbar\omega_p/k_BT$, $\Lambda=\hbar I_cn_0/(2e\omega_p)$, and
\begin{equation}
P(x,x')=\ln\left|\frac{\sin[\pi(x+x')/d]}{\sin[\pi(x-x')/d]}\right|\,.
\end{equation}

The action~\eqref{Spsi} is similar to the action of a single long Josephson junction~\cite{weEPL,KatoJPSJ}, but it contains a term non-local in the $x$-coordinate, instead of a local term proportional to $(\lambda_J\partial\psi/\partial x)^2$, where $\lambda_J$ is the Josephson penetration depth. The non-locality and the small pre-factor proportional to $\gamma s/D$ strongly reduce the (positive) contribution to the action from the $x$-derivatives of the phase difference. This favors the phase difference $\psi_0$ to be spatial inhomogeneous for stacks with a width $D$ exceeding several $\gamma s$. The tunneling occurs via creating a fluxon with the characteristic scale $x_0\sim n_0\gamma s$. The addition factor $n_0$, describing the number of junctions taking part in the tunneling process, increases both $x_0$ and $\Lambda$ which reduces the MQT rate. This effect is similar to the current-locking phenomenon~\cite{Mros,Goldobin}.

In further approximation, we represent $\psi_0(x,\tau)$ in the form $\psi_0(x,\tau)=F(x)q(\tau)$, where $F(x)$ is a trial function, describing the shape of the tunneling fluxon in the $x$-direction, and $q(\tau)$ plays a role of the collective coordinate.
The form of the function $F(x)$ is chosen from the physical reason that the fluxon nucleates at the junction's edge~\cite{wePRL,weEPL,wePRB}.
We considered different trial functions, and obtained similar results.
For all results in this paper, the function $F(x)$ used is
\begin{equation}
F(x)=\frac{C(x_0)}{1+(x/x_0)^2}\,,
\end{equation}
where $x_0$ is the characteristic size of the tunneling fluxon (see Fig.~\ref{FigJunction}), and the normalization constant $C (x_{0})$ is chosen such that $\int_0^ddxF^2(x)/d=1$. The analysis shows that $x_0$ is about several $\gamma s$ (for more details, see Ref.~\onlinecite{wePRB}).
Since for the LSCO IJJ stack under study $x_{0} \sim \gamma s \approx 20\,$nm $\ll D = 0.95\,\mu$m, the inhomogeneity of the phase difference in the $x$-direction is crucial to make a large difference from the conventional MQT theory.

\begin{figure}
\begin{center}
\includegraphics[width=0.48\textwidth]{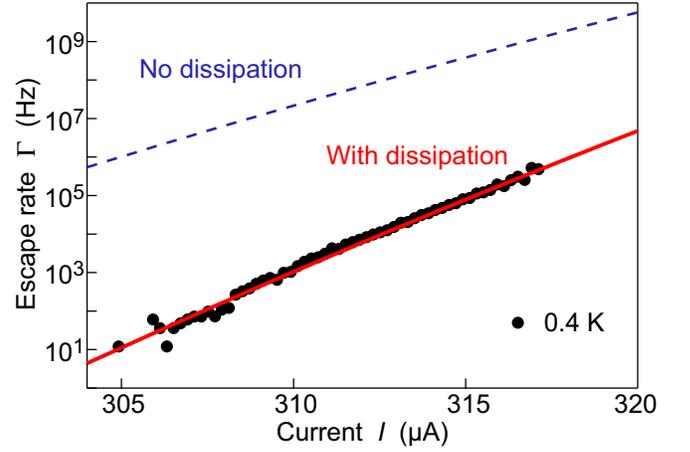}
\end{center}
\caption{\label{FigGamma}(Color online) The escape rate $\Gamma$ versus current $I$. Dots correspond to the experimental data at $T=0.4$\,K. The solid curve is the theoretical prediction. The parameters of the model are: $\gamma=30$, $n_0=5$, $s=0.7$\,nm, $D=0.95$\,$\mu$m, $I_c=344$\,$\mu$A, $f_p=440$\,GHz, $T=0$, and $\alpha=0.3$. The dashed curve is calculated for zero dissipation, $\alpha=0$. Other parameters are the same as for the solid curve.}
\end{figure}

Substituting $\psi_0(x,\tau)$ in the form $\psi_0(x,\tau)=F(x)q(\tau)$ into Eq.~\eqref{Spsi} and expanding the expression inside the integrals in a Taylor series, we obtain a particle-like action in the form:
\begin{equation}\label{Sq}
S_{\text{eff}}=\Lambda\!\!\int\limits_0^{\tau_0}\!\!d\tau\!\!\left[\frac{\dot{q}^2}{2}+U(q)+%
\frac{\alpha}{4\pi}\!\!\int\limits_{-\infty}^{\infty}\!\!\!d\tau'\frac{\left(q(\tau)-q(\tau')\right)^2}{(\tau-\tau')^2}\right]\!,
\end{equation}
where the potential $U(q)$ is
\begin{eqnarray}\label{U}
U(q)&=&\frac{\bar{\mu}q^2}{2}-\mu_0\sum_{n=2}^{\infty}\frac{(-1)^{n}\nu_{2n}}{(2n)!}q^{2n}\nonumber\\%
&&+j\sum_{n=1}^{\infty}\frac{(-1)^{n}\nu_{2n+1}}{(2n+1)!}q^{2n+1}\,.
\end{eqnarray}
In this equation,
\begin{equation}
\bar{\mu}=\mu_0+\frac{\mu_0n_0s}{4\pi\lambda_{ab} d^2}\!\!\int\limits_0^d\!\!dx\!\!\int\limits_0^d\!\!dx'%
\frac{\partial F(x)}{\partial x}P(x,x')\frac{\partial F(x')}{\partial x'}\,,
\end{equation}
and
\begin{equation}
\nu_n=\frac{1}{d}\int\limits_0^ddx\,F^n(x)\,.
\end{equation}
The analysis shows that for the parameters under study, we can neglect in Eq.~\eqref{U} the terms with $q^6$ and higher.
The potential $U(q)$ implicitly depends on the fitting parameter $x_0$.
The collective coordinate $q(\tau)$ in Eq.~\eqref{Sq} satisfies the imaginary-time equation of motion
\begin{equation}\label{EqQ}
\ddot{q}=\frac{\partial U}{\partial q}+\frac{\alpha}{\pi}\!\!\int\limits_{-\infty}^{\infty}\!\!\!d\tau'\frac{q(\tau)-q(\tau')}{(\tau-\tau')^2}
\end{equation}
with periodic boundary conditions $q(\tau+\tau_0)=q(\tau)$.
We solve this equation numerically for given $x_0$, and calculate the tunneling exponent $B=S_{\text{eff}}/\hbar$.
We repeat this procedure until finding the $x_0$ corresponding to the minimum of $B$.
The MQT escape rate is calculated then by Eq.~\eqref{Gamma0}.

At temperatures much smaller than the crossover temperature $T^{*}$, the escape rate $\Gamma$ is independent of $T$.
The experimental data for $\Gamma(I)$ calculated from $P(I)$, measured at $T=0.4$\,K, are shown in Fig.~\ref{FigGamma}. The results of  theoretical calculations for $\Gamma(I)$ at $T\ll T^{*}$, with the prefactor $a=1$, are also shown in this figure. There is a very good agreement between theory and experiment.
Here the dissipation parameter $\alpha = 0.3$, which is in reasonable agreement with the estimated value $0.2$, is found to be the best fit.
Note the strong dependence of the tunneling exponent $B$, and consequently of the MQT rate $\Gamma$ on the dissipation parameter $\alpha$.
Even a not too high dissipation constant $\alpha=0.3$ strongly reduces the escape rate $\Gamma$ (see the dashed and the solid curve in Fig.~\ref{FigGamma}).

\subsection{Thermal-activated regime}\label{SectionThermal}

Let us now focus on the thermal-activated regime.
At temperatures above the crossover temperature $T^{*}$, the solution of Eq.~\eqref{EqQ} with periodic boundary conditions $q(\tau+\tau_0)=q(\tau)$ becomes trivial: $q(\tau)=q_0$, where $q_0$ corresponds to the maximum of the potential $U(q)$.
The crossover temperature $T^{*}$ is defined as
\begin{equation}
T^{*}=\frac{\hbar\omega_p}{2\pi k_B}\left[\sqrt{|U^{''}_0|+\frac{\alpha^2}{4}}-\frac{\alpha}{2}\right]\,,
\end{equation}
where $U^{''}_0$ is the second derivative of $U(q)$ at $q=q_0$.
The theoretical prediction of $T^{*}$ turns out to be $\approx 2$\,K for the bias currents $I$ used in the measurement with the same parameters as in Fig.~\ref{FigGamma}, even though $T^{*}$ depends on the bias current $I$.
This prediction of $T^{*}$ is in good agreement with the experiments [see Fig. \ref{FigP}(b)].

\begin{figure}
\begin{center}
\includegraphics*[width=0.48\textwidth]{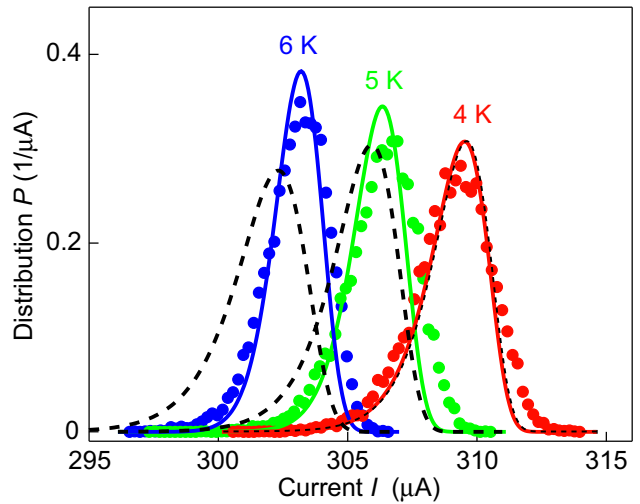}
\end{center}
\caption{\label{FigPT}(Color online) Switching current distributions $P(I)$ at three different temperatures above $T^{*}$.
Dots are experimental data.
Dashed curves are calculated according to Eq.~\eqref{P_Gamma}.
The parameters of the model are the same as for Fig.~\ref{FigGamma}. Solid curves are calculated taking into account the phase re-trapping effect.}
\end{figure}

For $T>T^{*}$, the tunneling exponent $B$ reduces to
\begin{equation}
B=\frac{\Lambda \omega_{p} U_0}{k_BT}\,,
\end{equation}
where $U_0=U(q_0)$ is the maximum of the potential $U(q)$.
To calculate the thermally activated escape rate,
 the value of $U_0$ has to be minimized with respect to the characteristic size of the fluxon $x_0$.
The tunneling exponent $B$ in the thermal-activated regime does not depend on the dissipation $\alpha$, and the escape rate~\eqref{Gamma0} only slightly depends on $\alpha$ due to the prefactor $a$. For $T>T^{*}$, the latter one can be written as~\cite{Caldeira,PreArev},
\begin{equation}
a=\sqrt{\frac{\bar{\mu}}{|U^{''}_0|}}\prod_{n=1}^{\infty}\left[\frac{\omega_n^2+\alpha\omega_n+\bar{\mu}}{\omega_n^2+\alpha\omega_n+U^{''}_0}\right],
\end{equation}
where $\omega_n=2\pi n/\tau_0$.

Now it is more convenient to consider the switching current distribution $P_{\mathrm{TA}}(I)$ in the thermal-activated regime instead of $\Gamma_{\mathrm{TA}}(I)$. Using Eq.~\eqref{Gamma_P}, we obtain
\begin{equation}\label{P_Gamma}
P_{\mathrm{TA}}(I)=\frac{\Gamma_{\mathrm{TA}}(I)}{(dI/dt)}\exp\left(-\frac{1}{(dI/dt)}\int\limits_0^{I}dI'\,\Gamma_{\mathrm{TA}}(I')\right)\,.
\end{equation}
The results of our calculations of $P_{\mathrm{TA}}(I)$ at three different temperatures $T>T^{*}$ are shown by the dashed lines in Fig.~\ref{FigPT}.
For all curves, the parameters are the same as for Fig.~\ref{FigGamma}. We see good agreement with experimental data at $T=4$\,K.
For larger temperatures, however, the peaks of the theoretical curves shift to lower currents, in comparison to the experimental curves.
In addition to that, the width of the theoretical $P_{\mathrm{TA}}(I)$ increases as the temperature increases, while the experimental  $P_{\mathrm{TA}}(I)$ shows the opposite behavior. We attribute these small discrepancies to the phase re-trapping (phase diffusion) process due to dissipation. In the next section, we will address and solve this issue.

\subsection{Thermal activation with phase re-trapping}\label{SectionRetrapping}

Thermal fluctuations stimulate the switching of the phase $\varphi$ of the junction to the running (resistive) state. However, similar fluctuations can also help re-trap $\varphi$ back to the metastable (superconducting) state. For more information about re-trapping in Josephson junctions, see, e.g., Ref.~\onlinecite{Krasnov}.

In contrast to the normal switching probability without dissipation in the thermal-activated regime $T>T{^{*}}$, the probability of re-trapping strongly depends on the dissipation parameter $\alpha$. We consider the re-trapping effect following the theory developed in Ref.~\onlinecite{Jacob}. There, a dissipative particle, moving in a periodic washboard potential is studied in the presence of thermal noise. The state of the particle with a constant average velocity corresponds to the resistive state of the Josephson junction. Thermal fluctuations can trap the particle in one of the local minima of the potential (corresponding to the superconducting state). The authors of Ref.~\onlinecite{Jacob} derived an analytical formula for the re-trapping rate $\Gamma_{r}$. Applying this to the Josephson junction, $\Gamma_{r}$ is then given by
\begin{equation}\label{GammaR}
\Gamma_r=\omega_p\sqrt{\frac{E_J(I-I_r)^2}{2\pi I_c^2k_BT}}\exp\left[-\frac{E_J(I-I_r)^2}{2I_c^2\alpha^2k_BT}\right],
\end{equation}
where $I_r=4\alpha I_c/\pi$ is the re-trapping current, and $E_J=\hbar I_c/(2e)$. This formula is valid for a single short Josephson junction. In our case the situation is much more complicated because the phase difference is inhomogeneous in the $x$-direction.
We assume that the phase difference can be ``partially re-trapped''. In this process, thermal fluctuations lead to the appearance of a ``re-trapped fluxon'' of width $x_r$ in the $x$-direction. After this, the junction dynamically relaxes to the metastable (superconducting) state. We modify Eq.~\eqref{GammaR} by multiplying the Josephson energy $E_J$ by the factor $x_r/D$, where we take $x_r$ equal to $x_0$.

Using Eq.~\eqref{GammaR} for the re-trapping rate (with an additional factor $x_0/D$ in $E_J$), we calculate the probability $F_{nR}(I)$ for the fluxon {\it not} being re-trapped. The relationship between $\Gamma_r(I)$ and $F_{nR}(I)$ becomes~\cite{Krasnov}:
\begin{equation}\label{Pr_Gamma}
F_{nR}(I)=\exp\left(-\frac{1}{(dI/dt)}\int\limits_{I}^{I_c}dI'\,\Gamma_r(I')\right)\,.
\end{equation}
Thus, the switching distribution, which is actually measured in the experiment, is equal to $P_{\mathrm{TA}}(I)F_{nR}(I)$, where $P_{\mathrm{TA}}(I)$ is given by Eq.~\eqref{P_Gamma}. The results of such calculations at three different temperatures, from $4$ to $6$ K, are shown in Fig.~\ref{FigPT} by the solid curves.
For all curves, the model parameters are the same as for Fig.~\ref{FigGamma}. Now we see that the theoretical curves show better agreement with the experimental data: the positions of $P(I)$ do not shift towards smaller currents, and the widths of the $P(I)$ decrease with increasing temperature.
For $T=4$\,K, the solid and dashed (no re-trapping) curves practically coincide.
This means that the re-trapping effect turns out to be not significant for $T\lesssim4$ K.
This correlates well with the temperature dependence of the standard deviation $\sigma$ [width of $P(I)$] in Fig.~\ref{FigP}(b). Namely, due to the existence of phase diffusion at $T\gtrsim4$\,K, $\sigma(T)$ shows unconventional temperature dependence, different from the $T^{2/3}$ law. This phenomenon [violation of the $T^{2/3}$ law of $\sigma(T)$], unique for the LSCO IJJ stack [cf. Refs.~\onlinecite{Inomata,Ustinov,Kashiwaya,KuboAPEX} (BSCCO) and Ref.~\onlinecite{Ueda} (HBCCO)], can be attributed to the relatively small superconducting gap of LSCO; namely, above 4 K there are still non-negligible quasi-particles which contribute dissipation to the IJJ stack, resulting in re-trapping.

\section{Conclusions}

We have measured the switching current distributions in a submicron LSCO IJJ stack for a wide temperature range, from $0.4$ K to $10$ K. The switching probability does not depend on temperature for $T\lesssim2$\,K, implying that the LSCO IJJ stack is in the MQT regime. We have applied and extended the theory for switching dynamics in IJJ stacks, taking into account the effect of dissipation both in the MQT and the thermal activated regimes. The theory and the experiments are in good agreement. We have also shown that dissipation plays an important role for all temperatures. The phase re-trapping takes place in the thermal-activated regime, where $T\gtrsim 4$\,K.

\section*{Acknowledgements}

The authors thank S. Kawabata, A. Tanaka, T.~Koyama, M.~Machida, I.~Kakeya, N.~Kobayashi, and Y.~Ootuka for discussions.
This work was partially supported by MEXT under Grant No. 1905014. YK was financially supported by NIMS. AOS acknowledges support from the Dynasty Foundation and the Russian Foundation for Basic Research (projects Nos. 12-02-92100 and 12-02-00339). FN is partially supported by the ARO, NSF grant No. 0726909, JSPS-RFBR contract No. 12-02-92100, Grant-in-Aid for Scientific Research (S), MEXT Kakenhi on Quantum Cybernetics,
and the JSPS via its FIRST program.

\appendix

\section{Determination of the fluctuation-free critical current}

\begin{figure}[t]
\begin{center}
\includegraphics[width=0.45\textwidth]{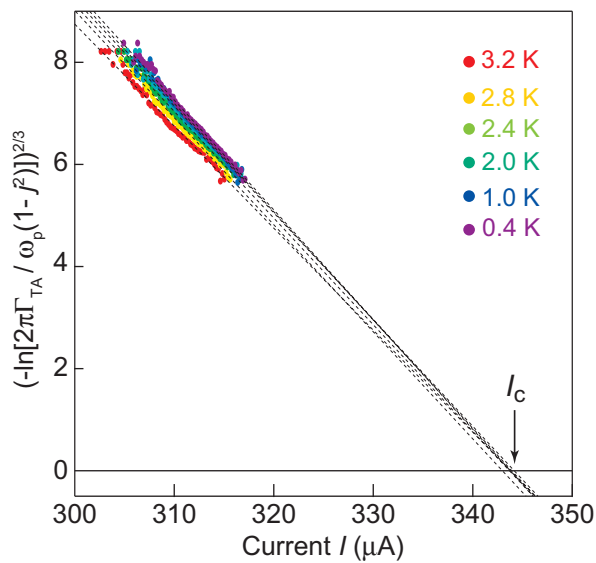}
\end{center}
\caption{\label{FigEscln}(Color online) Determination of $I_{c}$ from the experiments. Dots are experimental data converted using Eq.~\eqref{lnGamma}, and dashed lines are linear fits.
The mean value of $I$ at the intercepts of the extrapolated lines gives $I_{c} = 344$\,$\mu$A.}
\end{figure}

Here we describe how the fluctuation-free critical current $I_{c}$ was determined from the experiments.
We used the escape rate in the thermal-activated regime in the form
\begin{equation}
\label{GammaTA}
\Gamma_{\rm{TA}} = \frac{\omega_{p}}{2\pi}(1-j^2)^{\frac{1}{4}}\exp{\left( -\frac{\Delta U}{k_{B} T} \right)},
\end{equation}
where the prefactor $a$ is assumed to be unity.
Then Eq.~\eqref{GammaTA} is converted~\cite{Wallraff} to a linear function with respect to the bias current $I$,
\begin{equation}
\label{lnGamma}
\left( - \ln \left[ \frac{2\pi \Gamma_{\rm{TA}}}{\omega_{p}  (1-j^2)^{1/4} } \right] \right)^{\frac{2}{3}} =  \left( \frac{4\sqrt{2} E_{J}}{3 k_{B} T} \right)^{\frac{2}{3}} \frac{1}{I_{c}} (I_{c} - I),
\end{equation}
where the barrier height $\Delta U$ was approximated by a cubic function~\cite{Potential}: $\Delta U \approx \frac{4\sqrt{2}}{3} E_{J} (1-I/I_{c})^{3/2}$.
The normalized $\Gamma$ after this conversion is shown in Fig.~\ref{FigEscln}.
Here we used the following values as the initial guesses for the fit: $I_{c}= 340\,\mu$A, obtained by a rough fit with Eq.~\eqref{GammaTA} in which $I_{c}$ and $T$ are defined as free parameters, and $C = 135\,$fF determined from the geometry of the junction, i.e., $C = \varepsilon_{0}\varepsilon A/s$.

From the mean value of the zero-crossing of the extrapolated lines in Fig.~\ref{FigEscln}, $I_{c}$ was determined to be $344$\,$\mu$A.
Since $\omega_{p}$ is inside the logarithmic term on the left hand side of Eq.~\eqref{lnGamma}, the variation of the initial $I_{c}$ gives a very small difference to the result of the conversion~\cite{Wallraff} in Eq.~\eqref{lnGamma}.
Therefore it is sufficient to use the roughly estimated $I_{c}$ as an initial value, and consequently $I_{c}$ can be determined by repeating the fitting procedure with high accuracy.
The zero-bias plasma frequency was also obtained: $\omega_{p}= \sqrt{2e I_{c}/\hbar C} = 2\pi \times 440\,$GHz.
In Fig.~\ref{FigEscln}, the data obtained at temperatures $T\lesssim3.5$\,K have been chosen for the fit, because the re-trapping effect cannot be negligible at temperatures $T\gtrsim 4\,$K, as discussed in Sect.~\ref{SectionRetrapping} and shown in Fig.~\ref{FigPT}.

\begin{figure}[t]
\vspace{5mm}
\begin{center}
\includegraphics*[width=0.45\textwidth]{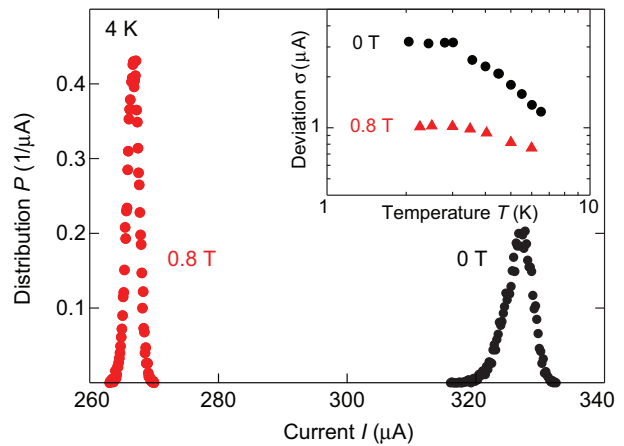}
\end{center}
\caption{\label{FigA2}(Color online) Switching current distributions $P(I)$ of another LSCO IJJ stack at 4 K under two different magnetic fields, 0 and 0.8 T. Inset: The standard deviation $\sigma$ as a function of temperature, at $0$ and $0.8$\,T. }
\end{figure}

\section{Effect of magnetic field}\label{Bfield}

Here we show that the critical current of another LSCO IJJ stack, which has a  geometry similar to the IJJ stack studied in this paper, is suppressed  under a magnetic field $H$ parallel to the \textit{ab}-plane ($z$-direction in Fig. \ref{FigJunction}).
As shown in Fig.\ref{FigA2}, the position of the switching current $P (I)$ under a magnetic field $H=0.8$\,T is much lower than for $0$\,T. Note that for this IJJ stack, the magnetic flux $\Phi$ with $H=0.8$\,T is much lower than the flux quantum $\Phi_{0}$ (which corresponds to $H=3$\,T for this IJJ stack). The temperature dependencies of the standard deviations $\sigma (T)$ at $H=0$ and $H=0.8$\,T are plotted in the inset of Fig. \ref{FigA2}. There one can see a consistent reduction of $\sigma (T)$ under a magnetic field, although for this sample $\sigma (T)$ was measured only down to $2$\,K.


\begin{thebibliography}{99}
\bibitem{Potential} Even though the potential of the LSCO IJJ stack is actually given by $U_{0}$ in the thermal regime as we discussed in sect. \ref{SectionThermal} [see also Eq.~\eqref{U}], here we instead use $\Delta U$ for simplicity of calculation since the both $U_{0}$ and $\Delta U$ are approximately proportional to $(1 - I/I_c)^{2/3}$; therefore both are equivalent in the following discussion.
\end{thebibliography}
\end{document}